\documentstyle[aps,multicol]{revtex}

\def\beginwide{
        \end{multicols} \vspace*{-0.5cm} \noindent
        \rule{3.5in}{.1mm}\rule{.1mm}{5mm} \widetext \medskip }
\def\endwide{
        \hspace*{3.35in}~\rule[-5mm]{.1mm}{5mm}\rule{3.5in}{.1mm}
        \begin{multicols}{2} \vspace*{-1.0cm} \noindent }
\tighten
\draft

\begin{document}

\title{Supression of magnetic subbands in semiconductor superlattices\\ 
driven by a laser f\,ield.}

\author{C. Rodr\'{\i}guez-Castellanos, M. T. P\'erez-Maldonado}
\address{Department of Theoretical Physics, Faculty of Physics\\ University of
Havana\\ San L\'azaro y L, Vedado, La Habana, Cuba, CP 10400\\
E-mail addresses: mtperez@ff.oc.uh.cu, crc@ff.oc.uh.cu}

\maketitle

\begin{abstract}
\noindent The effect of strong laser radiation on magnetic subbands in 
semiconductor superlattices is investigated. Due to the presence of a 
magnetic field perpendicular to the growth direction, non-linear effects 
such as band supression and electron localization become relevant at relatively 
lower intensities and for any polarization  
perpendicular to the magnetic field. Electron quasienergies and density of 
states are calculated in the Kramers-Henneberger approximation, whose 
validity is discussed. The conditions under which collapse of magnetic 
subbands and quenching of N-photon emission or absorption processes occur 
are discussed. We conclude that at laser frequencies close to cyclotronic 
frequency and intensities typical of c.w. lasers, magnetic subbands become flat, 
magnetotunneling is inhibited and multiphotonic processes dominate optical absorption.
\end{abstract}

\pacs{}

\begin{multicols}{2}

\section{Introduction.}

The continous increase of laser intensity is making possible the study of a
wide range of non-linear phenomena in atoms, plasmas and solids under the
action of intense electromagnetic radiation \cite{brandi}.

In the last decade these studies have been extended to semiconductor
nanostructures under intense electric fields, originated by an applied ac
voltage or a high-intensity infrared laser. At frequencies of the order of
$1\,THz$, typical of free-electron lasers, photon energies are comparable to
the energy separation of the electronic levels and nanostructures couple
strongly to the electromagnetic field. In the case of semiconductor
superlattices of period $d$, driven by electromagnetic radiation of frequency
$\omega $, amplitude $F$ and polarized in a direction perpendicular to the
interfaces, the I vs V curves are modified and non-linear behaviour is
observed. Holthaus \cite{holthaus1,holthaus2,holthaus3} has predicted the
collapse of minibands and the dynamical localization of electrons with the
co\-rres\-ponding supression of the coherent tunneling current if $F/F_0$ is
a zero of the Bessel function $J_0$, being $F_0=\hbar\omega/ed$. As a consequence of
Tucker's formula \cite{tucker}, tunneling processes in which the absorption or emission of 
N photons take place are inhibited when $F/F_0$ is a zero of $J_N$ . Sequential 
photon-assisted tunneling, with differential negative resistivity, dynamical electron 
localization, formation
of electric field domains and absolute negative conductance for low applied
voltages have been observed experimentally
\cite{tucker,drexler,keay1,keay2,zeuner1}. In the miniband transport regime 
the effect of inverse Bloch oscillations has been reported \cite{zeuner2}. 
Many of these phenomena have been theoretically explained 
\cite{unterrainer,platero,aguado} and several interesting predictions 
have been made: oscillations of the superlattice reflection coefficient 
of the radiation due to dynamical electron localization \cite{ghosh}, 
step-like behaviour of the interminiband multiphotonic absorption coefficient 
\cite{monozon}, spontaneous breaking of temporal symmetry with the 
appearance of a dc current in presence of an ac voltage \cite{alekseev} 
and the occurrence of Rabi oscillations when $\hbar\omega $ is close to 
the energy difference between two minibands \cite{diez}. A new kind of 
microwave frequency multiplier, an ultra-fast detector of $THz$ radiation 
and a Bloch laser based in the driven motion of the electrons in one 
miniband have been proposed \cite{renk}.

The energy spectrum of a superlattice of period $d$ under the action of a
magnetic field $B$ perpendicular to the growth direction has also been well
studied \cite{maan,melquiades}. Instead of Bloch minibands, one has Landau
magnetic subbands, with energies $E_n(x_0)+\hbar ^2k^2/(2m^*)$, where $x_0$,
($-d/2\leq x_0\leq d/2$) is the position of the centre of the cyclotronic
orbit, $k$ is the momentum along the magnetic field direction and $m^*$ the
conduction electron effective mass. Let $l_B=(\hbar c/eB)^{1/2}$ be the
magnetic length. If the ratio $d/l_B<1$, the first subbands are almost flat
and those of higher index $n$ present a dispersive character. If $d/l_B>1$,
all subbands are dispersive. The difference between these two regimes is
observed in optical intraband absorption at frequencies close to that of
cyclotronic resonance $\omega _c = eB/m^*c$ \cite{maan,duffield}. The
dispersion of magnetic subbands makes possible tunneling in the growth
direction, which occurs via transitions between consecutive subbands in the
anti-crossing points \cite{brey}.

This work considers a new, combined situation, in which a
superlattice undergoes the action of a magnetic field perpendicular to its
growth direction and intense electromagnetic radiation, linearly polarized in
a direction perpendicular to $\vec B$. We search for the occurrence of
effects such as the collapse of magnetic subbands and electronic localization
similar to those predicted for Bloch minibands. The magnetic field introduces
a new characteristic frequency (the cyclotron frequency) in the system under
consideration, and an enhanced effect of the external electromagnetic field
is expected near the resonance conditions. Additionally, $\vec B$ couples the
motion in the directions perpendicular to it and, therefore, a similar result
should be observed for waves polarized in either of these directions. We have
found that non-linear effects become important for relatively lower
intensities and radiation polarized in any direction perpendicular to the
magnetic field. Quasienergies and density of states in the Kramers -
Henneberger approximation are calculated and the conditions under which
collapse of magnetic subbands and supression of N-photon absorption or
emission processes occur are analyzed.

\section{Schrodinger equation. Kramers-Henneberger approximation.}

Let $OX$ be the growth direction of the superlattice and $OZ$ the magnetic
field direction. The electromagnetic radiation of frequency $\omega$ and
amplitude $F$ can be polarized either in $OX$ or $OY$ directions.

In the effective mass approximation, the envelope wavefunction of a
conduction electron can be written as
\beginwide
\begin{equation}
\label{envelopewf}
\psi (\vec r,t)=\frac{1}{S}\exp\left(-\frac{i}{\hbar}(qy+kz)\right)
\exp\left(-\frac{i}{\hbar}\frac{k^2}{2m^*}t\right)\hat U\varphi(x,t)
\end{equation}
\endwide
where $S$ is the transverse area of the sample, and $m^*$ is the conduction
electron effective mass. We have considered that $m^*$ is constant,
independent of $x$. The validity of this, commonly used, approximation has been
discussed in \cite{melquiades}. The time-dependent unitary transformation
$\hat U$ is given by \cite{galvao}:
\begin{equation}
\hat U=\exp\left(\frac{i}{\hbar}u(t)\hat p\right)\,
\exp\left(\frac{i}{\hbar}v(t)x\right)\,
\exp\left(\frac{i}{\hbar}w(t)\right)
\end{equation}
With the choice
%\beginwide
\begin{eqnarray}
\label{choice}
\nonumber u(t)&=&\frac{eF}{m^*}\left(\frac{\omega }{\omega _c}\right)^{s-1}
\frac{\cos\omega t}{\omega _c^2-\omega ^2}\\
v(t)&=&eF\omega _c\left(\frac{\omega _c}{\omega }\right)^s\frac{\sin\omega
t}{\omega _c^2-\omega ^2}\\
\nonumber w(t)&=&-\frac{e^2F^2}{4m^*(\omega ^2-\omega _c^2)}t-
\frac{e^2F^2(\omega _c^2+\omega ^2)}{8m^*\omega (\omega _c^2-\omega ^2)^2}
\sin 2\omega t\\
\nonumber &+&\frac{eFl_B^2q}{(\omega _c^2-\omega^2)}\omega _c
\left(\frac{\omega _c}{\omega }\right)^s\sin\omega t
\end{eqnarray}
%\endwide
for $\omega\neq\omega _c$, the function $\varphi (x,t)$ satisfies the
equation
\begin{equation}
\label{eqsch1}
i\hbar \frac{\partial\varphi}{\partial t}=\left(\hat H_0+\hat
W(t)\right)\varphi
\end{equation}
where
\beginwide
\begin{eqnarray}
&&\hat H_0=-\frac{\hbar ^2}{2m^*}\frac{d^2}{dx^2}+\frac{1}{2}m^*\omega
_c^2(x-x_0)^2+V_0(x) \\ &&\hat W(t)=2\sum_{N=1}^{\infty}V_N(x)\,\cos N\omega
t \\
\label{gral}
&&V_N(x)=V\sum_{m=-\infty}^{\infty}a_mJ_N\left(m\frac{F}{F_0}\right)
\,\exp\left(i\frac{2\pi mx}{d}\right)\;\;\;\;N=0,1,... \\
&&F_0(\omega )=\frac{edB^2}{2\pi m^*c^2}\left(\frac{\omega }{\omega
_c}\right)^s\left[\left(\frac{\omega }{\omega _c}\right)^2-1\right]\\
&&x_0=\frac{l_B^2}{\hbar}q
\end{eqnarray}
\endwide
Here $V\,a_m$ are the Fourier coefficients of the superlattice potential:
\begin{equation}
V(x)=V\sum _m a_m\,\exp\left(i\frac{2\pi mx}{d}\right)
\end{equation}
and $s$ takes values $0$ or $1$ for electromagnetic waves polarized along $OX$ or $OY$ axis.

The effect of radiation is expressed in the "dressed" potential $V_0(x)$,
which renormalizes magnetic subbands, and the superposition of harmonic terms
with amplitude $V_N(x)$, giving raise to N-photon emission and absorption
processes. The dynamical Stark effect is included in Eq.(\ref{choice}) for
$w(t)$ as a phase shift of the wavefunction, and is given by
\begin{equation}
\Delta E=\frac{e^2F^2}{4m^*(\omega ^2-\omega _c^2)}
\end{equation}
When $F<<F_0$, $V_N\sim V(F/F_0)^N$ and standard time-dependent perturbation
theory can be applied. Non-linear effects become important for $F\sim F_0$.
If $F>>F_0$, all $V_N$ have the same order of magnitude and $W\sim
V(F_0/F)^{1/2}$.

When $\omega $ is close to $\omega _c$, $F_0$ is relatively small. For
example, if $B=16\,T$, $m^*=0.067\,m_e$ and $d=20\,nm$, then $F_0\sim 10^2\,V/cm$.
This corresponds to intensities about $10^2\,W/cm^2$, which can be reached
with a c.w. laser. Therefore, the dynamics of recombination processes and
many-body effects due to high levels of photoexcitation need not to be
considered.

Quasienergies $\varepsilon $ are the eigenvalues of the operator $\hat
H_0+\hat W-i\hbar\partial /\partial t$ \cite{sambe} and can be obtained from
the poles of $\hat G(t)$, the retarded Green's function averaged over a
period $T=2\pi/\omega$ of the initial time. The Kramers-Henneberger
approximation (KHA), widely used in atoms \cite{volkova}, approximates
$\varepsilon$ by the eigenvalues of $\hat H_0$ and corresponds to the zeroth
order term of the expansion of $\hat G(t)$ in even powers of $\hat W$:
\begin{equation}
\label{expgreen}
\hat G(t)=\sum_{p=0}^{\infty}\langle\hat G_0\left(\hat W\hat
G_0\right)^{2p}\rangle
\end{equation}
where $\langle ...\rangle $ denotes the time-average operation described
above.  Following the arguments of \cite{antunes}, it is easy to show that
the above series converges absolutely. Note that $V_0$ enters the expansion
in zeroth order, whereas $V_N$ ($N>1$) contributes only to second and higher
orders. When $F>>F_0$ Eq.(\ref{expgreen}) is an expansion in powers of
$\alpha =(V/(\hbar\omega _c))(F_0/F)^{1/2}$. Then, as KHA neglects terms of
the order of $\alpha ^2$ it is expected to give a good result if $\alpha
^2<<1$. Since in this case $V_0(x)=Va_0+O(\alpha )$, the dressed potential
can also be considered an small perturbation partially breaking the
degeneracy of Landau levels.

It is important to realize that the limit $\alpha\to 0$ can be achieved not
only by increasing the laser intensity ($I\to\infty$) but also by approaching
the cyclotronic frequency ($\omega\to\omega _c$).

In what follows, we will consider a particular potential:
\begin{equation}
V(x)=\frac{V}{2}\left(1-\cos\frac{2\pi x}{d}\right)
\end{equation}
for which the expressions of $V_N(x)$ have the simple form
\begin{equation}
V_N(x)=\frac{V}{2}\left[\delta _{N,0}-J_N\left(\frac{F}{F_0}\right)
\cos\left(\frac{2\pi x}{d}+\frac{N\pi }{2}\right)\right]
\end{equation}
In this case, magnetic subbands collapse when $F/F_0$ is a zero of $J_0$, and
processes involving absorption or emission of N photons are quenched at zeros
of $J_N$.

However, for an arbitrary superlattice, $V_N(x)$ is given by Eqs.(\ref{gral})
and different harmonics are modified in different proportions, so $V_N(x)$
does not necessarily become zero for finite values of $F/F_0$. When
$F/F_0\to\infty $ ($\alpha\to 0$) and $N$ is fixed, $V_N$ approaches zero as
$V(F_0/F)^{1/2}$ and magnetic subbands are supressed. In this limit
multiphoton processes of many orders have a contribution of similar magnitude
to the absorption and emission probabilities.

To order $\alpha $ the quasienergies are given by the expression
%\beginwide 
\begin{equation}
\label{perturb}
E_n^{(1)}(x_0)=\hbar\omega
_c\left(n+\frac{1}{2}\right)+\frac{V}{2}-U_n\,\cos\frac{2\pi x_0}{d}
\end{equation}
%\endwide
where
%\beginwide
\begin{equation} 
U_n=\frac{V}{2}J_0\left(\frac{F}{F_0}\right)\,\exp\left(-\frac{\pi
^2l_B^2}{d^2}\right)L_n\left(\frac{2\pi ^2l_B^2}{d^2}\right)
\end{equation}
%\endwide
and $L_n(z)$ are Laguerre polynomials. 

\section{Results and discussion.}

Numerical diagonalization of $\hat H_0$ gives the electron quasienergies
$E_n(x_0)$ in the KHA, shown in Fig.1 (solid lines) for $V=0.3\,eV$,
$m^*=0.067\,m_e$, $B=16\,T$, $d=20\,nm$ and $F/F_0=0;5;10;15;20;30$,
corresponding to laser intensities lower than $10^5\,W/cm^2$ at frequencies 
$\omega\sim\omega _c$. We have also shown the quasienergies obtained from 
Eq.(\ref{perturb})(dashed lines).

The curves show the following features:
\begin{enumerate}
\item[a.] In the absence of radiation ($F=0$) all magnetic subbands are dispersive, 
in agreement with \cite{maan}, since for the values of $B$ and $d$ considered $d>l_B$. 
\item[b.] As $F/F_0$ increases, the magnetic subbands become flat.
\item[c.] As $F/F_0$ increases, the results derived from Eq.(\ref{perturb}) 
approach those calculated in KHA approximation.
\item[d.] The facts described in b. and c. do not occur monotonously, 
because of the oscillations of $J_0$. For example, these effects are stronger for 
$F/F_0=15$ than for $F/F_0=20$ because $J_0$ has a zero in $14.9309$. 
\item[e.] One must expect the same qualitative behaviour for
any superlattice, but in the general case there are no finite values of
$F/F_0$ for which magnetic subbands collapse. Note that in this case
\begin{equation}
V_0(x)=V\sum_{m=-\infty}^{\infty}a_mJ_0\left(m\frac{F}{F_0}\right)
\,\exp\left(i\frac{2\pi mx}{d}\right)
\end{equation}
\end{enumerate}
In Fig.2 we show the density of electron states (DOS) calculated from Eq.(\ref{perturb}) 
for $F/F_0=5$ and $15$. When magnetic subbands become flat, the time averaged velocity 
in the $OY$ direction is zero and electron motion becomes one-dimensional, as can be seen
in the shape of the DOS curve for $F/F_0=15$. Under these conditions, tunneling in the 
direction of growth is expected to be quenched, as for Bloch minibands.

We conclude that, when $\omega $ approaches $\omega _c$, magnetic subbands
are supressed, tunneling in the growth direction is inhibited and multiphoton
processes dominate intersubband absorption spectra.

\acknowledgments {This work has been partially supported by an Alma Mater Project of the
University of Havana. Authors thank Humberto S. Brandi for helpful
suggestions. One of us (C. R. C.) also acknowledges Carlos Tejedor and Gloria
Platero for warm hospitality and detailed discussions of the manuscript at
Universidad Aut\'onoma and Institute for Materials in Madrid, Spain.}

{\bf Figure captions:}

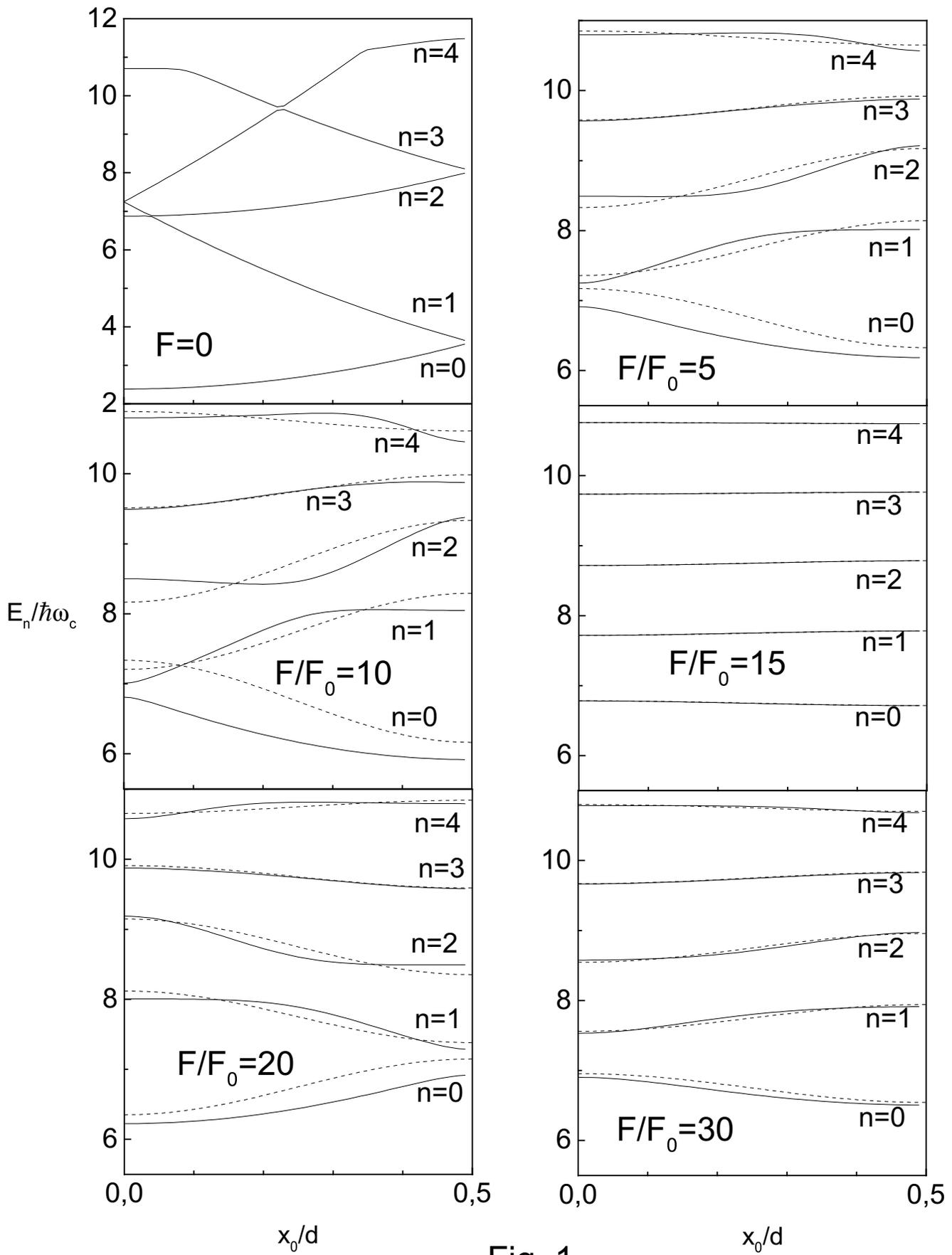
Fig 1. Electron quasienergies $E_n(x_0)$ ($n=0,1,2,3,4$) as functions of the
centre of the cyclotronic orbit for different values of $F/F_0$. (Solid lines:
KHA. Dashed lines: 1st order perturbation theory)

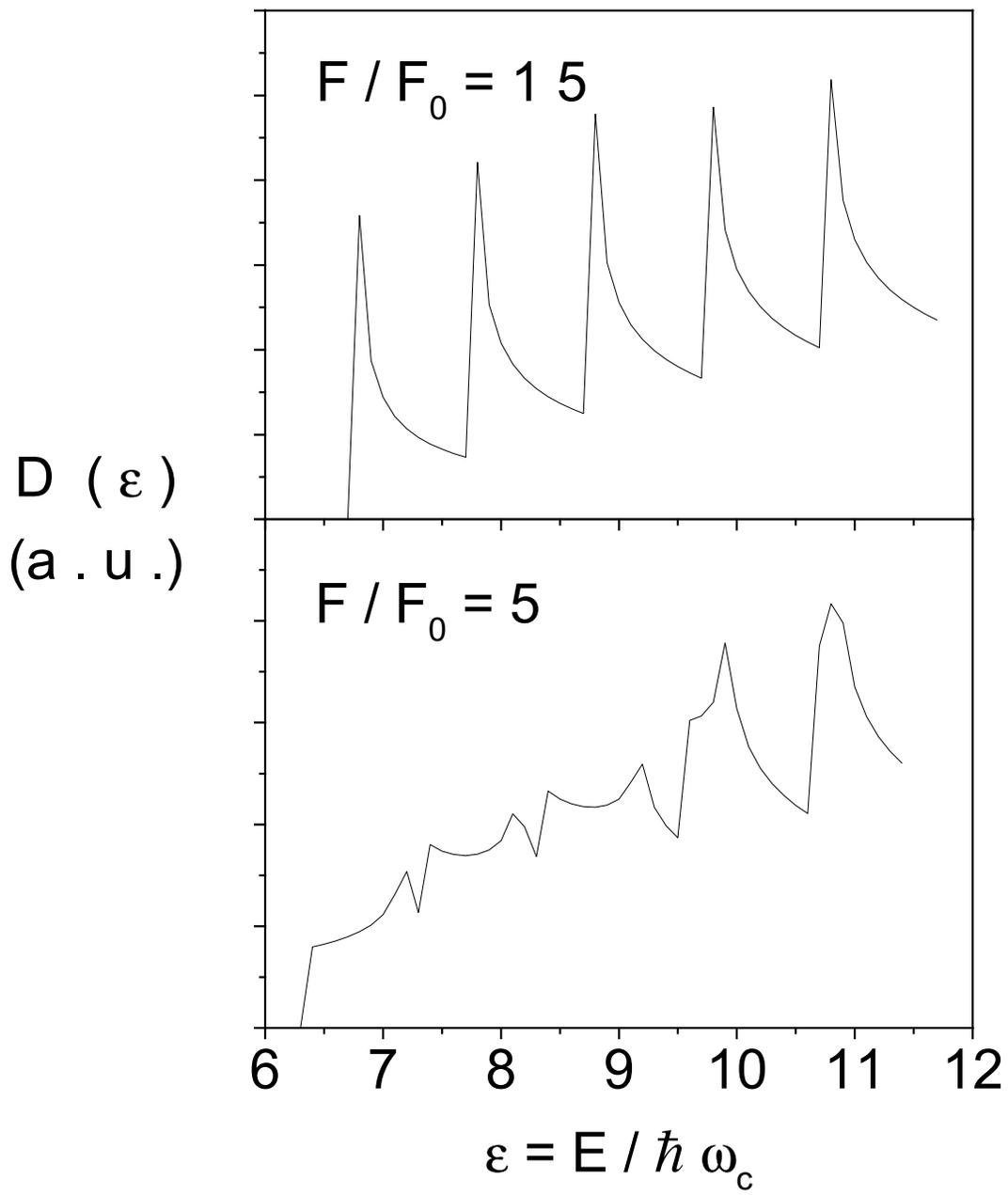
Fig 2. Density of electron states in 1st order perturbation theory for
$F/F_0=5$ and $15$.

\end{multicols}


\begin{thebibliography}{50}
 
\bibitem{brandi} H. S. Brandi, L. Davidovich, G. Jalbert, B. Koiller and 
N. Zagury, in {\it Intense Laser Phenomena and Related Subjects. Series of 
Optics and Photonics}, edited by I. Yu. Kyian and M. Yu. Ivanov, 
(World Scientific, Singapore, 1991), p. 299.

\bibitem{holthaus1} M. Holthaus, Phys. Rev. Lett. {\bf 69}, 351 (1992).

\bibitem{holthaus2} M. Holthaus, Z. Phys. Condensed Matter 
{\bf 89}, 251 (1992).

\bibitem{holthaus3} M. Holthaus and D. Hone, Phys. Rev. B 
{\bf 47}, 6499 (1993).

\bibitem{tucker} J. R. Tucker and M. J. Feldman, Rev. Mod. Phys. 
{\bf 57}, 1055 (1985).

\bibitem{drexler} H. Drexler, J. S. Scott, S. J. Allen, Jr., K. L. Campman 
and A. C. Gossard, Appl. Phys. Lett. {\bf 67}, 2816 (1995).

\bibitem{keay1} B. J. Keay, S. J. Allen, Jr., J. Galan, J. P. Kaminsky, 
K. L. Campman, A. C. Gossard, U. Bhattacharya and M. J. W. Rodwell, 
Phys. Rev. Lett. {\bf 75}, 4098 (1995).

\bibitem{keay2} B. J. Keay, S. Zeuner, S. J. Allen, Jr., K. D. Maranowsky, 
A. C. Gossard, U. Bhattacharya and M. J. W. Rodwell, Phys. Rev. Lett. 
{\bf 75}, 4102 (1995).

\bibitem{zeuner1} S. Zeuner, B. J. Keay, S. J. Allen, Jr., K. D. Maranowsky, 
A. C. Gossard, U. Bhattacharya and M. J. W. Rodwell, Phys. Rev. B 
{\bf 53}, R1717 (1996).

\bibitem{zeuner2} S. Zeuner, S. J. Allen, Jr., K. D. Maranowsky and A. C. Gossard, Appl. Phys. Lett. {\bf 69}, 2689 (1996).

\bibitem{unterrainer} K. Unterrainer, B. J. Keay, M. C. Waanke, S. J. Allen, Jr., D. J. Leonard, G. Medeiros Ribeiro, U. Bhattacharya and M. J. W. Rodwell , Phys. Rev. Lett. {\bf 76}, 2973 (1996).

\bibitem{platero} G. Platero and R. Aguado, Appl. Phys. Lett. {\bf 70},
3546 (1997).

\bibitem{aguado} R. Aguado and G. Platero, Phys. Rev. Lett. 
{\bf 81}, 4971 (1998).

\bibitem{ghosh} A. W. Ghosh, A. V. Kuznetsov and J. W. Wilkins, 
Phys. Rev. Lett. {\bf 79}, 3494 (1997).

\bibitem{monozon} B. S. Monozon and A. G. Zilich, Phys. Solid State 
{\bf 37}, 508 (1995).

\bibitem{alekseev} K. N. Alekseev, E. H. Cannon, J. C. McKinney, 
F. V. Kusmartsev and D. C. Campbell, Physica D {\bf 113}, 129 (1998).

\bibitem{diez} E. Diez, R. G\'omez-Alcal\'a, F. Dom\'{\i}nguez-Adame, 
A. S\'anchez and G. P. Berman, Phys. Lett. A {\bf 240}, 109 (1998).

\bibitem{renk} K. F. Renk, E. Schomburg, A. A. Ignatov, J. Grenzer, 
S. Winnerl and K. Hofbeck, Physica B {\bf 244}, 196 (1998).

\bibitem{maan} J. K. Maan, Festk\"orperproblem {\bf 27}, 137 (1987).

\bibitem{melquiades} M. de Dios Leyva and V. Galindo, Phys. Rev. B 
{\bf 48}, 4516 (1993).

\bibitem{duffield} T. Duffield, R. Bhat, M. Koza, F. DeRosa, K. M. Rush 
and S. J. Allen, Jr., Phys. Rev. Lett. {\bf 59}, 2693 (1987).

\bibitem{brey} L. Brey, G. Platero and C. Tejedor, Phys. Rev. B 
{\bf 38}, 9649 (1988).

\bibitem{galvao} R. M. O. Galvao and L. C. M. Miranda, Am. J. Phys. 
{\bf 51}, 729 (1980).

\bibitem{sambe} H. Sambe, Phys. Rev. A {\bf 7}, 2203 (1973).

\bibitem{volkova} E. A. Volkova, A. M. Popov and O. V. Smirnova, JETP 
{\bf 79}, 736 (1994).

\bibitem{antunes} H. S. Antunes-Neto and L. Davidovich, 
Phys. Rev. Lett. {\bf 53}, 2184 (1987).

\end{thebibliography}
\end{document}